\documentclass[12pt]{article}
\usepackage{epsfig} 
\usepackage{amssymb}
\catcode`\ä = \active 
\catcode`\ü = \active 
\catcode`\ö = \active 
\catcode`\Ä = \active 
\catcode`\Ü = \active 
\catcode`\Ö = \active 
\catcode`\ß = \active
\letß=\ss 
\defä{\"a} 
\defü{\"u} 
\defö{\"o} 
\defÖ{\"O} 
\defÄ{\"A}
\defÜ{\"U} 

\def\IJMP#1{Int. J. Mod. Phys.~{\bf #1}}

\def\NP#1{Nucl. Phys.~{\bf #1}}

\def\PL#1{Phys. Lett.~{\bf #1}}
\def\PR#1{Phys. Rev.~{\bf #1}}

\def\PRT#1{Phys. Rept.~{\bf #1}}

\def\ZP#1{Z. Phys.~{\bf #1}}

\begin{document}

\title{The Phase Diagram of the Quark-Meson Model}

\author{B.-J. Schaefer$^a$  and J. Wambach$^{a,b}$\\[2mm]
  {\small $^a$Institut f\"{u}r Kernphysik, TU Darmstadt, D-64289
    Darmstadt, Germany}\\
  {\small $^b$Gesellschaft f\"{u}r Schwerionenforschung GSI, D-64291
    Darmstadt, Germany} }


\maketitle

\noindent{PACS: }

\begin{abstract} 
  
  Within the proper-time renormalization group approach, the chiral
  phase diagram of a two-flavor quark-meson model is studied. In the
  chiral limit, the location of the tricritical point which is linked
  to a Gaussian fixed point, is determined. For quark chemical
  potentials smaller than the tricritical one the second-order phase
  transition belongs to the $O(4)$ universality class. For
  temperatures below the tricritical one we find initially a weak
  first-order phase transition which is commonly seen in model studies
  and also in recent lattice simulations.  In addition, below
  temperatures of $T \lesssim 17$ MeV we find two phase transitions.  The
  chiral restoration transition is initially also of first-order but
  turns into a second-order transition again.  This leads to the
  possibility that there may be a "second tricritical'' point in the
  QCD phase diagram in the chiral limit.

\end{abstract}

\newpage

\section{Introduction}

Recently, the phase diagram of the chiral phase transition has
received renewed interest with the emergence of improved lattice
techniques at finite quark chemical potential, $\mu$. Through
heavy-ion collision experiments at RHIC, the LHC and the future GSI
facility parts of the phase diagram may be accessible.  As one of the
prominent features of the QCD phase diagram, the existence of a
(tri-)critical endpoint of a first-order phase transition line has
emerged.  This is strongly suggested by model calculations and also
indicated by lattice simulations. The precise location of this point
is still unknown. In the chiral limit, the critical endpoint becomes a
tricritical one corresponding to a Gaussian fixed point. In this
limit, the chiral finite-temperature transition at vanishing $\mu$ is
likely to be of second-order for two-flavor QCD, where the critical
transition line is expected to fall into the $O(4)$-universality class
in three dimensions~\cite{pisarski}. For finite current masses, the
$O(4)$-transition line converts to a smooth crossover and the
tricritical point turns into a fixed point belonging to the $3d$ Ising
universality class~\cite{stephanov,hatta,raja}.

Second-order phase transitions are characterized by long-wavelength
fluctuations of the order parameter. Thus, the critical and
non-critical regions of the phase diagram, including first-order phase
transitions, can best be described by renormalization group (RG)
methods. We employ an analytical RG approach, the proper-time
renormalization group (PTRG) \cite{liao,dario,litim}, to investigate
the QCD phase diagram for two quark flavors. As a result of
integrating the quantum and thermal fluctuations with a certain cutoff
scale $k$, we obtain a flow equation for the scale-dependent action.
Following the scale evolution towards $k=0$, we arrive at the
effective action which contains all thermodynamically relevant
universal and non-universal information.

The paper is organized as follow: In the Sec.~\ref{secii} we derive
the analytical PTRG flow equation for an optimized smearing function.
This allows to find analytical threshold functions for finite
temperatures and chemical potentials. In this way, the numerical
summation of Matsubara terms can be circumvented and the numerical
error especially for small temperatures during the evolution is
reduced. In Sec.~\ref{seciia} we consider analytical limits of the
full flow equation in order to elucidate its intuitive structure and
sketch the numerical implementation in Sec.~\ref{seciib}. The
resulting phase diagram for massless quarks is presented and explored
in some detail in Sec.~\ref{seciii}. Finally, we summarize and
conclude in Sec.~\ref{conclusion}.

\section{The flow equations for finite $T$ and $\mu$}
\label{secii}

As an effective realization of two-flavor QCD, we use the effective
action of a linear quark-meson model defined at a scale $\Lambda$ in
the ultraviolet (UV) region of the theory in four Euclidean dimensions
by\footnote{The argument $\Phi$ of the effective action is a 
  short-hand notation of all incorporated fields.}
\begin{equation}
\label{effaction}
\Gamma_\Lambda [\Phi] =\! \int\!\! d^4 x \left\{ \bar q [\gamma^\mu
  \partial_\mu \!+\! 
  g (\sigma \!+\! i\vec{\tau}\vec{\pi}\gamma_5)] q + \frac 1 2
  (\partial_\mu \sigma)^2 + \frac 1 2 (\partial_\mu \vec{\pi})^2 +
  V_\Lambda (
  \sigma^2 \!+\! \vec{\pi}^2)  \right\}
\end{equation} 
with one isoscalar-scalar $\sigma$-meson and three
isovector-pseudoscalar pions $\vec{\pi}$.  The fermionic fields $q(x)$
and $\bar q(x)$ incorporate $N_f = 2$ flavor and $N_c =3$ color
degrees of freedom.  $\vec{\tau}$ labels the three Pauli matrices.
The Yukawa coupling $g$ is kept constant during the RG evolution.  The
purely mesonic general effective potential term $V_\Lambda$ is a
function of the $O(4)$-symmetric field $\vec{\phi}^2 = \sigma^2 \!+\!
\vec{\pi}^2$. As initial condition in the UV we choose for $V_\Lambda$
the truncation
\begin{equation}
V_\Lambda (\sigma^2 \!+\!\vec{\pi}^2) = \frac{m^2_\Lambda} 2 (\sigma^2 \!+\!
\vec{\pi}^2) + \frac{\lambda_\Lambda} 4 (\sigma^2 \!+\!
\vec{\pi}^2)^2
\end{equation} 
with the parameters $m^2_\Lambda$ and $\lambda_\Lambda$ to be
specified later. The wavefunction renormalizations for the meson- and
quark fields are neglected in this approximation.

The coarse-grained, infrared-finite and renormalization-group improved
flow equation for the scale dependent effective action $\Gamma_k
[\Phi]$ of a given theory is governed by
\begin{equation}\label{mastereq}
\partial_t \Gamma_k [ \Phi ] = -\frac 1 2 \int\limits_0^\infty
\frac{d\tau}{\tau} \left[ \partial_t f_k (\tau
k^2) \right] \mbox{Tr}  \exp \left( -\tau \Gamma^{(2)}_k [\Phi]
\right)\ ,
\end{equation} 
where Tr denotes a four-dimensional momentum integration and a trace
over all given inner spaces (e.g.~Dirac, color and/or flavor-space).
Details concerning the derivation of this flow equation and inherent
approximations can be found in Refs.~\cite{liao,litim1}.
$\Gamma^{(2)}_k [\Phi]$ represents the full inverse propagator and is
given by a second functional derivative of $\Gamma_k$ with respect to the
appropriate field components
\begin{equation} 
\Gamma^{(2)}_k [\Phi] = \frac{\delta^2 \Gamma_k
  [\Phi]}{\delta\Phi \delta \Phi}
\end{equation} 

If one replaces the effective action derivative $\Gamma^{(2)}$ on the
right-hand side of Eq.~(\ref{mastereq}) by the corresponding classical
action $S^{(2)}$ one obtains the one-loop effective
action~\cite{liao,litim}. The replacement of the classical action
$S^{(2)}$ by the scale-dependent effective action on the right-hand
side of Eq.~(\ref{mastereq}) is called RG improvement.

By means of a Schwinger proper-time regularization, an infrared-finite
renormalization group equation (PTRG) is obtained. The infrared (IR)
scale is given by $t = \ln \left( k/\Lambda\right)$ where $\Lambda$
denotes the high momentum (UV) scale. One important ingredient of the
flow equation (\ref{mastereq}) is the $a$ $priori$ unknown smearing or
blocking function $f_k (\tau k^2)$ which regulates the UV and IR
divergent proper-time integration, respectively.

The convergence of the approximate solution of the flow equation is
partly controlled by the regulator and can be accelerated if a certain
blocking function is chosen.  In previous works~\cite{bohr,papp} the
regulator was given by
\begin{equation}
f_k^{(i,d)} = \frac {2^i
  (d-2)!!}{\Gamma (d/2)(d-2+2i)!!} \Gamma (d/2+i,\tau k^2)
\end{equation} 
for $d=4$ dimensions.  In Refs.~\cite{bohr,papp} details concerning
the choice of regulators and the scheme dependence of the proper-time
flow equations for the quark-meson model can be found.

This choice has the disadvantage that the generalization to
finite temperature results in Matsubara sums which have to be
evaluated numerically. For higher smearing functions i.e.~for
functions with indices $i>1$ more and more additional terms in the
flow equations are generated at finite $T$ and $\mu$ and render the
flow equations more complicated. The choice of the same regulator but
now for $i=1$ and in $d=3$ dimensions, produces an analytic RG flow
equation in $T$ as well as in $\mu$ and is given explicitly by
\begin{equation}\label{smearingfkt3.0}
\partial_t f_k^{(i=1,d=3)} (\tau k^2) = -\frac{8}{3 \sqrt{\pi}}
\left( \tau k^2 \right)^{5/2} e^{-\tau k^2}\ .
\end{equation}
Then only one threshold function for each physical degree of freedom
arises and the structure of the full RG flow equation for the
thermodynamic potential in the vacuum becomes simple and physically
intuitive. Also the finite $T$ and $\mu$ generalization is
straightforward.

In thermal equilibrium, we apply the Matsubara technique and follow
the standard procedure. The quark chemical potential is introduced in
the fermionic piece of the Lagrangian by the replacement $\partial_0
\to \partial_0 - i\mu$ in the time-component of the derivative.  The
finite-temperature extension of this PTRG method is described in
detail in Ref.~\cite{bohr} for an $O(N)$-model and in Ref.~\cite{papp}
for the quark-meson model.

In the Matsubara formalism, the integration over the zero-component of
the momentum is replaced by a summation over Matsubara frequencies,
$\omega_n$, for mesons and, $\nu_n$, for the quarks:
\begin{eqnarray}  
\omega_n = 2n\pi T&\ , \ &\nu_n = (2n+1) \pi T\ ,\qquad n \in Z\!\!\!Z\ ,
\end{eqnarray} 
such that for an arbitrary function $g(p)$
\begin{equation}
\int \frac{d^4p}{(2\pi)^4} g(\vec p, p_4) \to T\sum\limits_{n=-\infty}^\infty
\int \frac{d^3p}{(2\pi)^3} g(\vec p, \left\{ \begin{array}{c} \nu_n + i\mu \\
  \omega_n  \end{array}\right\} )
\end{equation}
for Fermions and Bosons, respectively.

In order to solve the RG equation (\ref{mastereq}), a truncation of
the effective action is necessary. We will use the lowest-order in a
derivative expansion of the effective action given in
Eq.~(\ref{effaction}) at a given UV scale $\Lambda$.  This yields a
single partial differential equation for the scale-dependent grand
canonical thermodynamic potential $\Omega (T,\mu; \phi)$ which depends
on $T$, $\mu$ and the fields $\phi$:
\begin{eqnarray} 
\partial_t \Omega_k (T,\mu; \phi) &=& -\frac T 2 \int\limits_0^\infty
\frac{d\tau}{\tau} \left[ \partial_t f^{(i=1,d=3)}_k \right] 
\sum_n\!\! \int \frac{d^3 q}{(2\pi)^3} \mbox{Tr} \left[  
  e^{-\tau(\omega^2_n+\vec{q}^2+ \frac {\partial^2 \Omega_k (T,\mu;
      \phi)}{\partial \phi_i 
      \partial \phi_j} )} \right. \nonumber \\
&&\hspace*{3cm}\left. - e^{-\tau \left[ (\nu_n +i\mu)^2+ 
    \vec{q}^2 + g^2 \phi^2\right]}\right]  \ .
\end{eqnarray} 
Only the fermionic potential contribution is modified explicitly by
the finite quark chemical potential. At this stage, one sees nicely
the additive separation of the mesonic and fermionic flow in this
approximation.  Note, that we use the same blocking function for the
bosonic and fermionic flow-equation contributions.

Evaluating the traces and performing the remaining three-momentum
integrals, we finally obtain
\begin{eqnarray} 
&&\partial_t  \Omega_k (T,\mu; \phi)  = -\frac T {2(4\pi)^{3/2}}
\int\limits_0^\infty 
\frac{d\tau}{\tau^{5/2}} \left[ \partial_t f_k^{(i=1,d=3)} \right] \\
&& \sum_n \left\{ 
3 e^{-\tau(\omega^2_n+ 2 \Omega'_k )}+
    e^{-\tau(\omega^2_n+ 2 \Omega'_k + 4\phi^2 \Omega''_k )}- 4 N_c N_f
e^{-\tau \left[ (\nu_n +i\mu)^2+ g^2 \phi^2\right]} \right\}\nonumber 
\end{eqnarray} 
where the $\phi$-dependence of the potential derivatives on the
right-hand side has been suppressed. The primed potential denotes the
$\phi^2$-derivative, {\it i.e.} $\Omega'_k := {\partial \Omega_k}/
{\partial \phi^2}$ etc.  In the following, we define the pion-,
$\sigma$-meson and quark energies as
\begin{equation} 
E_\pi = \sqrt{k^2 + 2 \Omega'_k}\ ,\ 
E_\sigma = \sqrt{k^2 + 2 \Omega'_k + 4\phi^2 \Omega''_k}\ , \ 
E_q = \sqrt{k^2 + g^2 \phi^2}\ .
\end{equation} 
Evaluating the potential $\Omega_k$ at the global minimum, $\phi_0$,
the scale-dependent effective meson masses are defined as
$m_{\sigma,k}^2 = (2 \Omega'_k + 4\phi^2 \Omega''_k) |_{\phi=\phi_0} =
4\phi_0^2 \Omega''_k(\phi_0)$ and $m_{\pi,k}^2 = 2 \Omega'_k
|_{\phi=\phi_0}=0$. The dynamically generated (constituent) quark
masses $m_{q,k} = g \phi_0$ are proportional to the minimum since we
do not consider the running of the Yukawa coupling constant.

Finally, the PTRG flow equation for the scale-dependent grand
canonical potential can be integrated analytically, resulting in
\begin{eqnarray} 
\label{fullRG}
\partial_t \Omega_k(T,\mu; \phi) &=& \frac {k^5} {12\pi^2} 
\left[  \frac 3 {E_\pi} \coth \left(\frac {E_\pi}{2T} \right) +
    \frac 1  {E_\sigma} \coth \left(\frac
      {E_\sigma}{2T} \right) \right.\nonumber \\ 
&& \hspace{4ex}\left.- \frac {2 N_c N_f}{E_q}\left\{ \tanh
      \left(\frac 
     {E_q -\mu} {2T}\right) +\tanh \left(\frac
     {E_q +\mu} {2T}\right)\right\}\right]\ .
\end{eqnarray} 
This flow equation is the full RG equation in the approximation
considered for finite temperature and chemical potential. Note, that
since we have neglected the running of the Yukawa coupling, the
approximation does not correspond to the local potential approximation
(LPA).  It is possible to rewrite Eq.~(\ref{fullRG}) by means of
occupation numbers (cf.~Ref.~\cite{braun,meyer}).

The different degrees of freedom contribute in an additive way to the
flow. One recognizes the three (degenerate) pion-, the sigma- and the
quark/anti\-quark threshold functions in the square brackets. The flow
equation has an overall scale factor $k^4$, the correct dimension of
the potential in $d=4$ dimensions, which can be seen explicitly by
rewriting all threshold functions in a dimensionless form. Due to the
fermion loops, the fermionic contributions enter with a negative sign
and have a degeneracy factor of $(2s+1)\times N_c\times N_f$ with
$s=1/2$. Only in the quark/antiquark threshold functions the quark
chemical potential enters with the appropriate sign as it should be.

\subsection{Analytical limits}
\label{seciia}

It is instructive to investigate analytical limits of the full flow
equation (\ref{fullRG}). The flow in the vacuum is given by
\begin{eqnarray}
\label{zeroRG}
\partial_t \Omega_k (0,0; \phi) &=& \frac {k^5} {12\pi^2} \left[  \frac 3
  {{E}_\pi} + 
    \frac 1  {{E}_\sigma} - \frac {4 N_c N_f}{{E}_q}\right] \ .
\end{eqnarray}
Due to spontaneous chiral symmetry breaking, the quark energies $E_q$
are always positive and contribute to the flow.  In the mesonic
threshold functions poles can emerge, depending on the shape of the
potential. For instance, the pion pole for $\phi=0$ is determined
exactly by the condition $\Omega'_k = -k^2/2$.  However, for an
appropriate choice of initial conditions at the UV scale $\Lambda$,
the poles are never reached during the $k$-evolution.

For finite $T$ but vanishing $\mu$, the difference between fermionic
and bosonic Matsubara sums in $\partial_t \Omega_k (T,0)$ vanishes in
the low-temperature limit and we regain the vacuum flow analytically.
For finite $\mu$ and $T=0$, on the other hand, the flow is governed by
the non-analytic equation
\begin{eqnarray}
\label{muRG}
\partial_t \Omega_k (0,\mu; \phi) &=& \frac {k^5} {12\pi^2} \left[  \frac 3
  {{E}_\pi} + 
    \frac 1  {{E}_\sigma} - \frac {4 N_c N_f}{{E}_q}
    \left\{ 1 -  \theta ({\mu} - {E}_q ) \right\} \right]
\end{eqnarray}
reflecting the existence of the Fermi surface.  Thus, in the
zero-temperature limit, the tanh-expressions in the fermionic
threshold functions of Eq.~(\ref{fullRG}) degenerate to $2 \theta (
{E}_q - {\mu} )$.  It is also possible to analytically perform the
zero chemical-potential limit of Eq.~(\ref{fullRG}) which again yields
the correct vacuum flow since the fermionic flow contribution in the
curly brackets tends to two.

For hadronic matter in thermal equilibrium, two new energy scales $T$
and $\mu$ enter the evolution equation. The quark chemical potential
influences the bosonic part $\partial_t \Omega_{k}^{
\mbox{\footnotesize Boson}}$ only implicitly through the meson masses.
If the bosonic fluctuations are neglected by setting $\partial_t
\Omega_k^{\mbox{\footnotesize Boson}} =0$, one obtains (for a constant
Yukawa coupling) standard mean-field theory results for the
thermodynamic potential. The remaining quark contribution can then
easily be integrated~\cite{mocsy}. This feature is also seen in other
RG approaches (e.g.~in Ref.~\cite{wetterich}). As is well known,
mean-field theory does not always give a reliable description of the
phase transition near criticality as it neglects the role of the
important (thermodynamic) fluctuations. We will come back to this
point below.

Rewriting the step-function as in Eq.~(\ref{muRG}), one recognizes
easily that the quark flow contribution splits into a vacuum and
finite-chemical potential part with opposite sign. For $\mu < E_q = g
\phi_{0,k=0}$, the finite-$\mu$ part vanishes and there is no
distinction to the vacuum flow. In the vacuum, spontaneous chiral
symmetry breaking is mainly driven by the quark fluctuations, while
meson fluctuations tend to restore chiral symmetry due to the relative
sign in the flow Eq.~(\ref{fullRG}). The relative sign between the
vacuum and finite density part in the quark sector of Eq.~(\ref{muRG})
suppresses the quark fluctuations which are important for chiral
symmetry restoration at high densities.

\subsection{Solving the flow equation}
\label{seciib}

In order to solve the flow Eq.~(\ref{fullRG}) numerically one has in
principle two possibilities: either one can discretize the unknown
potential $\Omega_k$ on a $\phi^2$-grid or expand the potential in
powers of $\phi^2$ around its minimum $\phi_0$.  The advantage of a
potential expansion is that only a few coupled flow equations have to
be solved, depending on the number of couplings in the potential. But
for each higher order of the potential expansion, a new coupled beta
function arises which increases the numerical effort drastically.  A
further disadvantage is that the potential, after the evolution, is
only known at the (local) minimum $\phi_0$.

The big advantage of the grid solution is that one knows the potential
not only for the minimum but also for arbitrary $\phi^2$. This is of
importance for a first-order phase transition where two degenerate
minima of the potential emerge. In order to describe the first-order
transition correctly, the knowledge of all local minima is required.
This is cumbersome in a potential expansion, except for some simple
potentials. For finite pion masses, i.e.~with an additional explicit
symmetry breaking term in the potential, every minimum has always a
finite value and the symmetry is never exactly restored.  Also in this
case, a precise determination of the critical temperature of a
first-order transition is very difficult within an expansion scheme
around only one minimum.  The solution on a grid is, however,
computationally very demanding since one encounters highly-coupled and
rather large matrix equations in order to obtain reliable accuracy.
Nonetheless, for the reasons given above, we opted for the grid
solution but will also compare the findings with results from a
potential expansion.

In order to solve the flow equation (\ref{fullRG}), we discretize the
field $\phi^2$ for a general potential term on a regular grid. Details
of the algorithm which has been adapted to include fermions, can be
found in Ref.~\cite{bohr} and references therein.  The task is then to
solve the resulting closed coupled set of flow equations for the
potential by starting the evolution in the ultraviolet and integrate
towards $k = 0$ for each grid point. In this limit, all quantum- and
thermal fluctuations are taken into account. It turns out that the
numerical convergence is faster if one works with the dimensionful
flow equation instead of the rescaled one.  The initial conditions are
chosen at the UV scale $\Lambda$ in such a way that they match
physical vacuum quantities obtained in the IR.  Predictions for finite
temperature and density are then possible without any further
parameter adjustments.  As initial condition we use the tree-level
parameterization of the symmetric potential
\begin{eqnarray}
\label{vinitial}
\left. V_\Lambda (\phi^2) = \frac{\lambda}{4} (\phi^2)^2
-\frac{\lambda}{2} \phi_0^2 \phi^2 \right|_{\Lambda}
\end{eqnarray}
at the compositeness scale $\Lambda = k_{\chi SB}$. For the results
presented below we choose $\Lambda = 500$ MeV, $\lambda = 10$ and
$\phi_0^2 = 0$.  The number of grid points is typically chosen around
$60 - 100$ for a $\phi$-interval between $0$ and $100$. We have
tested, that the numerical results do not change by varying the number
of grid points. The precise value of $\phi_0^2$ is not important,
since it has little influence on the vacuum results in the infrared
where chiral symmetry is spontaneously broken.  The Yukawa coupling is
fixed to $g = 3.2$ in order to reproduce a constituent quark mass of
the order of $300$ MeV.  These choices of initial parameters yield a
vacuum pion decay constant of $f_\pi \sim 87$ MeV in the chiral limit
which is consistent with values obtained from chiral perturbation
theory~\cite{chiralpt} and previous works~\cite{bohr}. For finite and
realistic pion masses, this value is shifted to $f_\pi \sim 93$ MeV.

\section{The phase diagram for $N_f = 2$}
\label{seciii}

Recently, significant progress has been made in studies of the QCD
thermodynamics with chemical potential by numerical simulations of
lattice gauge theory. Different methods of extrapolation to the
finite-$\mu$ region are in reasonable agreement
\cite{allton,owe,fodor} concerning the pseudocritical line $T_c(\mu)$
out to chemical potential values of several $100$
MeV~\cite{katz,karsch03,kars}.  This allows for a direct comparison of
lattice simulations with our nonperturbative method.  Some recent
results of the leading non-trivial term of quadratic order in the
Taylor expansion of the pseudocritical line are listed in
Tab.~\ref{tabcurv} and compared to the one obtained with the PTRG
method.

\begin{table}[!htb]
\begin{center}
\begin{tabular}{l|l}
    & \rule[-7mm]{0mm}{6ex} $\displaystyle T_c \left.\frac{d^2
        T_c}{d\mu^2}\right|_{\mu = 0 }$ \\ \hline 
\rule[-1mm]{0mm}{3ex}PTRG (full)        &  $\displaystyle -0.23$ \\ 
\rule[-1mm]{0mm}{3ex}PTRG ($\phi^4$)    &  $\displaystyle -0.258$ \\ 
\rule[-1mm]{0mm}{3ex}Lattice (Taylor)   &  $\displaystyle -0.14(6)$ \\
\rule[-1mm]{0mm}{3ex}Lattice ($\mu_I$)  &  $\displaystyle -0.101$ \\ 
\rule[-1mm]{0mm}{3ex}NJL (mean-field)   &  $\displaystyle   -0.4$ \\ 
\end{tabular}
\caption{\label{tabcurv} The curvature of the $N_f=2$ transition line
  for small quark chemical potentials compared with different
  methods.}
\end{center}
\end{table}

The curvature of the RG without any truncation, labeled by PTRG
(full), is smaller compared to a $\phi^4$ truncation of the potential
within the same RG approach, denoted by PTRG ($\phi^4$).  At the
expected tricritical point, a first-order phase transition emerges
where higher monomials in the potential expansion will become
necessary.  One the other hand, we know that the truncation works well
for the finite temperature and zero chemical potential phase
transition. It turns out that the $\phi^4$ truncation breaks down at
finite quark chemical potentials around $\mu \sim 100$
MeV~\cite{bjjwwork}.

Note that the RG results are obtained for zero quark masses in
contrast to the lattice simulations (see e.g. table 1 in
Ref.~\cite{ishikawa} for recent lattice pion masses in physical
units). On the lattice (e.g.~Ref.~\cite{allton}), realistic light
quark masses have not yet been achieved in a Taylor series estimate of
the reweighting factor up to quadratic order in $\mu$. The first
Taylor coefficient corresponds to the curvature and is labeled by
(Taylor) in the Table.  The equivalent coefficient obtained with the
imaginary chemical potential method in denoted by ($\mu_i$) in the
Table~\cite{owe}.  However, the lattice results of Ref.~\cite{allton}
suggest that any dependencies of the curvature on the quark masses are
weak. Compared to the chiral limit we see a difference of a factor
two.  A mean-field calculation within a NJL-model in the chiral limit
yields the largest curvature.  However, the curvature depends strongly
on the chosen parameter set for the NJL model~\cite{buballahab}.

\begin{figure}[!htb]
  \centerline{\hbox{ 
      \psfig{file=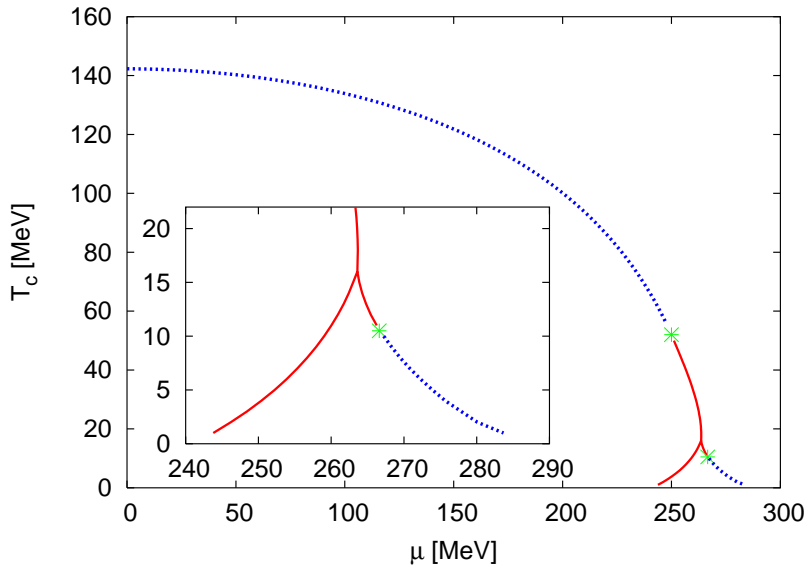,width=0.8\textwidth,angle=0}}}
\caption{\label{figphasedia} The phase diagram of the two-flavor 
  quark-meson model in the chiral limit. The tricritical points are
  denoted by an asterisk. For $\mu$ below the upper tricritical point
  the transition is of second-order (dashed line) and belongs to the
  $O(4)$ universality class. For temperatures smaller than $T<17$ MeV
  the single first-order phase transition (solid line) splits into two
  phase transitions.  The left transition line is always of
  first-order while the chiral restoration transition line around the
  splitting point is initially of first-order (solid line) and turns
  into a second-order (dashed line) phase transition for smaller
  temperatures.}
\end{figure}

In Fig.~\ref{figphasedia} we show the chiral phase diagram for the two
flavor linear quark-meson model in the chiral limit. At $\mu = 0$, a
second-order phase transition with a critical temperature of $T_c \sim
142$ MeV is found in which the spontaneously broken chiral symmetry is
restored.  Note that the critical temperature is not a universal
quantity and depends on the fitted pion decay constant in the vacuum.
The phase transition and fixed point structure is in the $O(4)$
universality class and was investigated within the same PTRG framework
in detail in Ref.~\cite{bohr}.

For realistic pion masses the second-order phase transition is washed
out and becomes a smooth crossover. This results in a larger
``critical'' temperature of the order of $180$ MeV~\cite{bj99}.

For finite chemical potential, the second-order phase transition with
$O(4)$ critical exponents persists up to a tricritical point. The
location of this tricritical point is found to be at $T_{c, tri} = 52$
MeV and $\mu_{c, tri} = 251$ MeV for the initial conditions chosen.
Within a nonperturbative expansion scheme, similar results are found
in Ref.~\cite{patkos}. This point belongs to a trivial Gaussian fixed
point with mean-field critical exponents as we have verified
explicitly~\cite{tetradis}.

For finite (realistic) pion masses, model calculations and lattice
simulations suggest a range of $T \sim 100 -180$ MeV and $\mu_B \sim
50 -700$ MeV for the location of the critical point of
QCD~\cite{stephanov}. The critical mean-field region of the endpoint,
where the mean-field theory of phase transitions is still valid, is
further expected to be small~\cite{hatta}.  Nevertheless, the
qualitative features of the phase diagram which we address, do not
depend on the chosen initial conditions. Thus, the existence of this
point, the shape of the transition lines and its universality class is
a prediction within the underlying model.

\begin{figure}[!htb]
  \centerline{\hbox{
      \psfig{file=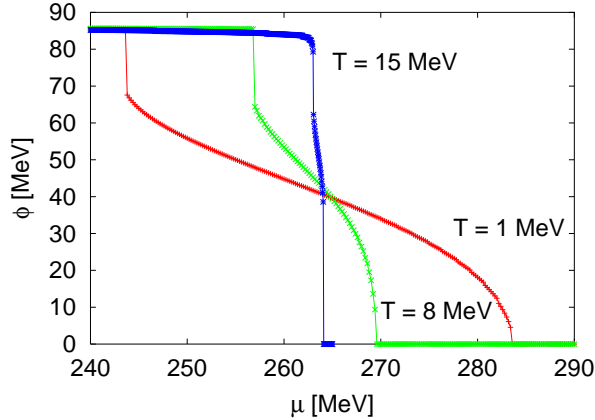,width=0.6\textwidth,angle=0}}}
\caption{\label{figVEV} The order parameter $\phi$ versus chemical potential 
  for three different temperatures.}
\end{figure}

At higher chemical potentials and smaller temperatures, the phase
transition changes initially to a single first-order phase transition.
For critical temperatures below $T_{s} \sim 17$ MeV and around $\mu_s
\sim 263$ MeV, however, we observe a splitting of the transition line
and two phase transitions emerge. The left transition line (see
magnified panel in Fig.~\ref{figphasedia}) represents a first-order
transition down to the $T=0$-axis. At this transition, the order
parameter jumps not to zero but to a finite value (Fig.~\ref{figVEV}).
The chiral symmetry remains spontaneously broken and is only restored
for higher chemical potentials which produce the second (right)
transition line.  For temperatures just below the splitting point,
i.e.~for $T<17$ MeV, at the right transition line we initially find a
second first-order transition where the order parameter again jumps to
zero and chiral symmetry gets restored.  But for smaller temperatures
close to the $\mu$-axis the order parameter tends smoothly to zero.
Due to a finite grid spacing it is difficult for such small
temperatures to determine the precise order of the transition. It
seems, however, that it is again of second-order. The determination of
the corresponding universality class is postponed to future
work~\cite{bjjwwork}.  Nevertheless, we infer that there must be a
second tricritical point in the phase diagram.

In Fig.~\ref{figVEV} the order parameter, which corresponds to the
expectation value (VEV) of the $\sigma$-field, is shown as function of
the chemical potential for three different temperatures in the
splitting area of the phase diagram (Fig.~\ref{figphasedia}). Around
the splitting point at $T = 15$ MeV, the order parameter depends
slightly on the chemical potential $\mu$ before the first transition,
thus washing out the edge at the gap. The $\mu$-dependence becomes
weaker for decreasing temperatures. For $T=1$ MeV there is almost no
difference between the finite-density and vacuum evolution and the
order parameter stays at the vacuum value $\phi \sim 87$ MeV.  For
$T=15$ MeV two gaps in the order parameter corresponding to two phase
transitions can be seen. The magnitude of the first gap is almost
constant and independent of the temperature.  For smaller temperatures
($T < 8$ MeV), the VEV as a function of $\mu$, smoothly decreases to
zero after the first transition and stays zero for larger chemical
potentials.  This is a signal that the system undergoes a second-order
transition again.  Finally, chiral symmetry is completely restored. It
is interesting to observe that all curves intersect roughly at the
same critical chemical potential of the splitting point, $\mu_s$,
where the magnitude of the VEV is about one half of the vacuum value.

The exact location of the splitting point and the first-order gap of
the order parameter depends on the chosen vacuum quark masses. For
larger vacuum quark masses e.g.~of the order of $380$ MeV,
corresponding to a larger Yukawa-coupling of $4.2$, the location of
the splitting point moves down to smaller temperatures and larger
chemical potentials. The gap in the order parameter of the first-order
transition increases with the quark masses.  Thus, varying the vacuum
quark masses within say $250-450$ MeV, the area bounded by the
transition lines is reduced. Yet, the order of the phase transition
lines are not altered.

\begin{figure}[!htb]
  \centerline{\hbox{
      \psfig{file=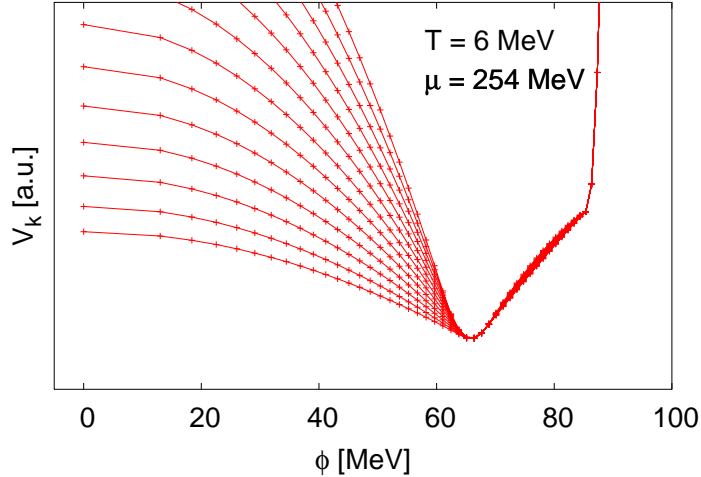,width=0.7\textwidth,angle=0}}}
\caption{\label{figpotb} The scale evolution of the thermodynamic
  potential towards the infrared is shown as a function of the field
  $\phi$ below the splitting point for $T=6$ MeV and $\mu = 254$ MeV.
  (Lowest line $k=8$ MeV,$\Delta k = 1$ MeV between each line). Since
  the chemical potential is larger than the critical one, both minima
  of the potential (one at $\phi \sim 65$ MeV and the other one at
  $\phi \sim 87$ MeV) are no longer equal. }
\end{figure}

\begin{figure}[!htb]
  \centerline{\hbox{
      \psfig{file=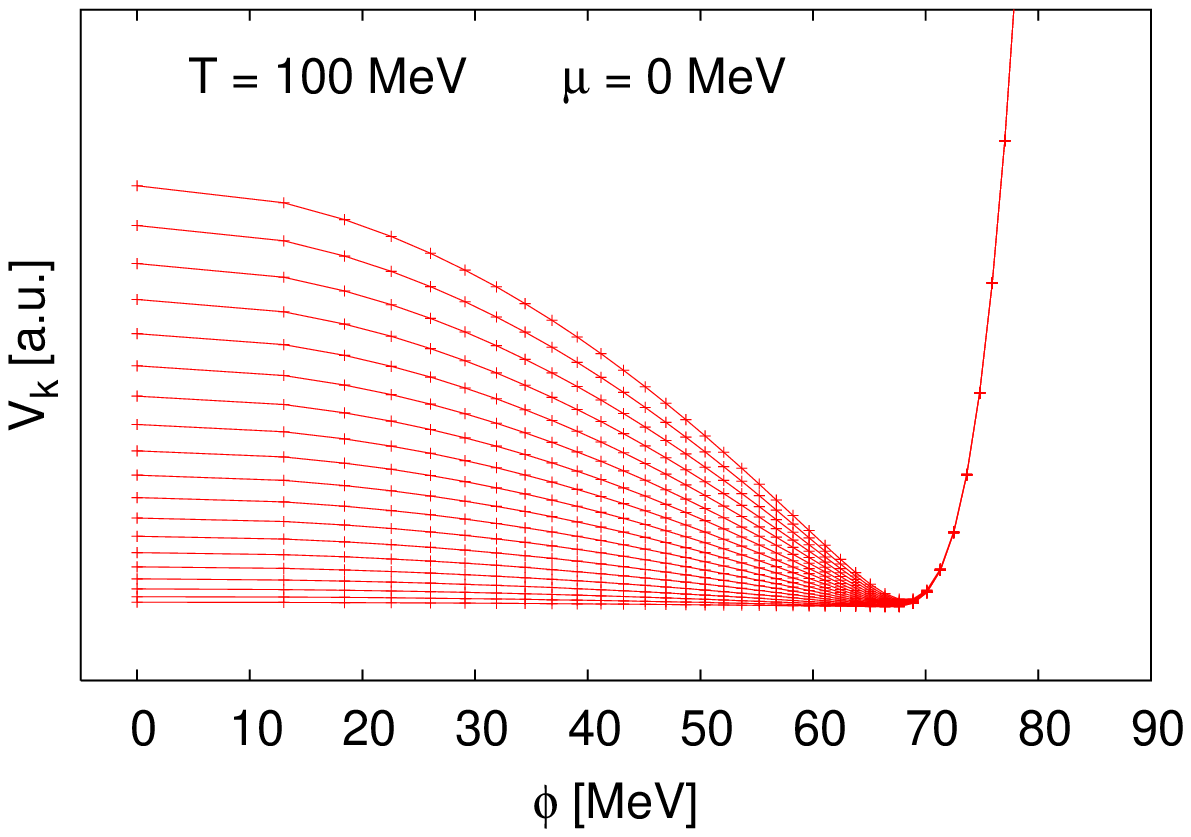,width=0.55\textwidth,angle=0}
      \hspace*{-4ex}
      \psfig{file=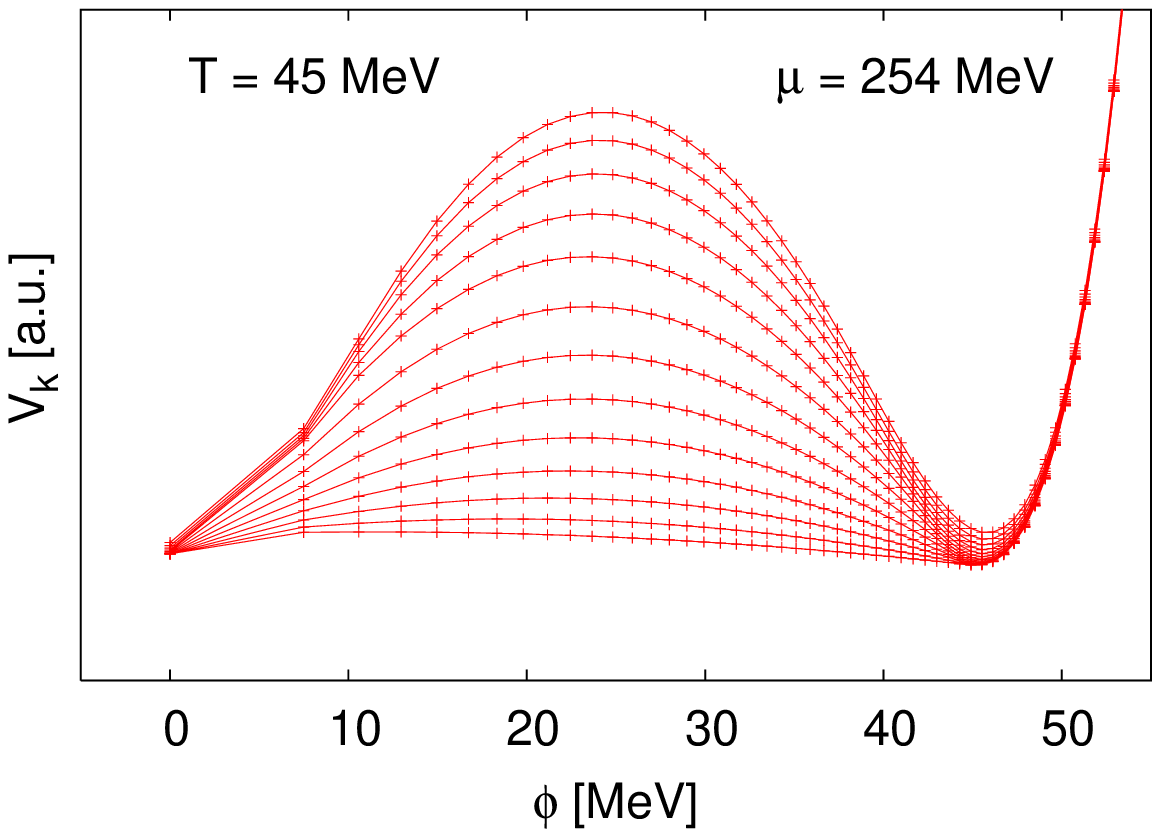,width=0.55\textwidth,angle=0}
    }}
    \caption{Typical scale evolution towards the infrared of a 
      second-order phase transition (left panel) and first-order
      transition (right panel). (Lowest line $k=2$ MeV, $\Delta k = 1$
MeV between each line. See text for further details).}
    \label{figpota}
\end{figure}

It is also instructive to see both phase transitions in the
thermodynamic potential explicitly. In Fig.~\ref{figpotb} the
$k$-evolution below the splitting point is shown for $T=9$ MeV and
$\mu = 254$ MeV for different (small) IR scales.
The potential exhibits two finite minima.\footnote{For any finite IR
  scale (which is numerically always realized) the potential is not
  exactly convex. Only in the IR limit $k=0$, it becomes convex as can
  be seen in the Fig.~\ref{figpota} \cite{bonanno}.}  This situation
is definitely different compared to that in Fig.~\ref{figpota}.  Here
the scale evolution of a typical second- or first-order transition is
shown in the left- and right panel, respectively.
For a second-order phase transition we have only one (global) minimum.
In the IR, the potential becomes flat (constant) below the minimum and
is almost scale independent above . The scale evolution barely effects
the minimum.  For the chosen initial conditions, this situation is
encountered for chemical potentials below $250$ MeV and temperatures
between $142$ - $52$ MeV.  A typical first-order transition is shown
in the right panel of Fig.~\ref{figpota} near the critical point.
Here, a second minimum at the origin emerges and the global minimum
jumps to zero for larger chemical potentials.  In all cases, the
potential tends to a convex function but the scale evolution of the
nontrivial minimum is very weak. This justifies to terminate the
numerical evolution at a finite but small IR scale (typically $k<1 -
10$ MeV)~\cite{bonanno}.

The pressure $p(T,\mu) = - \Omega_{k=0}(T,\mu,\phi_0)$ evaluated at
the global minimum $\phi_0$ is normalized to zero in the vacuum. For
finite chemical potential and vanishing temperature, it stays strictly
zero up to the transition point. Therefore, the quark number density,
$n_q = {\partial p}/{\partial \mu}$, jumps to a finite value.
Afterwards it increases monotonically.  Quantitatively, the quark
number density jumps to $0.1$ fm$^{-3}$ at $\mu \sim 242$ MeV which
corresponds to a normalized baryon density $\rho_B/\rho_0 \sim 0.18$.
These values are obtained for a vacuum quark mass of $\sim 275$ MeV
and increase significantly towards more realistic values if the vacuum
quark mass is increased.  These observations strongly suggest that the
system undergoes a gas-liquid transition to bound quark matter, where
the chiral symmetry remains spontaneously broken.  Thus, the
``triangular region'' in Fig.~\ref{figphasedia} consists of a bound
mixture of massive quarks/antiquarks, interacting massless pions and
massive sigma mesons.

\section{Conclusions}
\label{conclusion}

In order to provide a reliable identification of the phase structure
of strongly interacting matter, one must invoke nonperturbative
methods such as the Wilsonian formulation of the renormalization
group. We have employed a self-consistent RG approach with a smooth
proper-time regularization.  As an effective realization of QCD for
two flavors, a linear quark-meson model has been used with an
arbitrary $O(4)$-symmetric mesonic potential term. To explore the
phase diagram in the $(T,\mu)$-plane we have extended previous studies
to finite temperature and quark chemical potential. The resulting RG
equations for the potential term have been solved numerically in the
chiral limit.  The first non-trivial coefficient of the transition
line which belongs to the $O(4)$ universality class, has been
determined and compared with recent lattice simulations. The curvature
in the chiral limit is larger by a factor two as compared to lattice
analyses. The latter, however, employ unrealistically large $u-d$
current quark masses.

The location of the tricritical point, governed by a Gaussian fixed
point, has been determined. This structure of the phase diagram is
similar to the results of other works, where the second-order
transition line ends in a tricritical point and turns into a single
phase transition of (weak) first-order for decreasing
temperatures~\cite{tetradis,patkos}. This first-order transition
persists up to zero temperature and finite chemical potential.

For $\mu>\mu_c$ we also find initially a first-order transition which
splits into two phase transitions for temperatures below $17$ MeV. The
left transition line is always of first-order. The value for the
splitting point depends on the initial condition for dynamically
generated constituent quark mass in the vacuum. Apart from the
dependencies on the initial conditions, further uncertainties arise
from the truncation scheme of the flow equation which may affect the
exact values of the critical temperature and chemical potential. It is
difficult to obtain quantitative error estimates in this context. The
main focus of this work is not a quantitative prediction of some
non-universal numbers but the qualitative feature of the phase diagram
of the two-flavor linear quark-meson model.

For larger quark masses the splitting point moves towards smaller
temperatures but all qualitative features of the transition lines
persist.  Along the chiral symmetry restoration line below the
splitting point, the transition is initially also of first-order but
then turns into a second-order transition. This leads to the
suggestion that the QCD phase diagram may have a second
``tricritical'' point.

The triangular region in the phase diagram which is a new feature of
our calculation, is not related to spinodal curves or metastable
phases.  It is an effect driven by fluctuations. In a mean-field
calculation, we only find one single first-order transition in which
chiral symmetry is restored (see also~\cite{buballa}). The full RG
treatment, however, renders the order parameter and hence the
constituent quark mass finite beyond the first-order transition line.
At the transition the quark-number density jumps from zero to a finite
value and increases monotonically with increasing $\mu$.  We
interprete this transition as the liquid-gas transition to a state of
bound quark matter, consisting of interacting constituent quarks.

There are several directions in which the present analysis should be
extended. A obvious one is to include explicit chiral symmetry
breaking via finite current quark masses~\cite{bjjwwork}. This will
turn the second-order transitions into smooth cross overs. From
lattice studies it is well known that the strange quarks mass plays a
decisive role in the location of the critical endpoint. For infinitely
heavy strange quarks, in fact, it lies at $\mu=0$! Thus, it is
mandatory to extend our calculations to $N_f=3$. While this is a
computational challenge it will provide more realistic results. Work
in this direction is in progress.

\section*{Acknowledgments}

We would like to thank Michael Buballa for useful and illuminating
discussions. This work was supported in part by BMBF grant 06 DA 116.


\end{document}